\documentclass[12pt,preprint]{aastex}
\usepackage{graphics,graphicx,rotating,amsmath}

\newcommand{\lr}[1]{\left( #1 \right)}
\newcommand{\lrs}[1]{\left[ #1 \right]}
\newcommand{\bbe}{{\boldsymbol \beta}}
\newcommand{\bth}{{\boldsymbol \theta}}
\newcommand{\bde}{{\boldsymbol \delta}}
\newcommand{\bep}{{\boldsymbol \epsilon}}
\newcommand{\bepZ}{{\bep_{0th}}}

\newcommand{\mbe}{{\mbox{\boldmath$e$}}}
\newcommand{\mbeZ}{{\mbe_{0th}}}
\newcommand{\mbg}{{\mbox{\boldmath$g$}}}
\newcommand{\mbk}{{\mbox{\boldmath$k$}}}

\newcommand{\cZ}{{\cal Z}}

\newcommand{\hatA}{{\hat A}}
\newcommand{\hatD}{{\hat D}}
\newcommand{\hatG}{{\hat G}}
\newcommand{\hatI}{{\hat I}}
\newcommand{\hatP}{{\hat P}}
\newcommand{\hatS}{{\hat S}}
\newcommand{\hatR}{{\hat R}}

\begin{document}
\title{The systematic error test for PSF correction in weak gravitational lensing shear measurement by The ERA Method by Idealizing PSF}
\author{Yuki Okura\altaffilmark{1}} 
\email{yuki.okura@riken.jp}

\author{Toshifumi Futamase\altaffilmark{2}}
\email{}

\altaffiltext{1}
 {RIKEN Nishina Center}
\altaffiltext{2}
 {Department of Astrophysics and Atmospheric Science, Kyoto Sangyo University,}

\begin{abstract}
We improve the ERA(Ellipticity of Re-smeared Artificial image) method of PSF(Point Spread Function) correction in weak lensing shear analysis in order to treat realistic shape of galaxies and PSF. This is done by re-smearing PSF and the observed galaxy image smeared by a RSF(Re-Smearing Function), and allows us to use a new PSF with a simple shape and to correct PSF effect without any approximations and assumptions. We perform numerical test to show that the method applied for galaxies and PSF with some complicated shapes can correct PSF effect with systematic error less than 0.1\%. We also apply ERA method for real data of Abell 1689 cluster to confirm that it is able to detect the systematic weak lensing shear pattern. The ERA method requires less than 0.1 or 1 second to correct PSF for each object in numerical test and real data analysis, respectively.
\end{abstract}

\section{Introduction}
It is now widely recognized that weak gravitational lensing is an unique and powerful tool to obtain mass distribution in the universe. Coherent deformation of the shapes of background galaxies carries not only the information of intervening mass distribution but also the  cosmological background geometry and thus the cosmological parameters(Mellier 1999, Schneider 2006, Munshi et al. 2008). 

In fact weak lensing studies have revealed the averaged mass profile for galaxy cluster (Okabe et al. 2013, Umetsu et al. 2014) and detected the cosmic shear that is weak lensing by large scale structure is expected to be useful for studying the property of dark energy.   However the signal of cosmic shear is very weak and difficult to get useful constraint on the dark energy.   
Currently, several surveys are just started and planned to measure the cosmic shear accurately enough to constrain the dark energy property, such as Hyper Suprime-Cam on Subaru (http://www.naoj.org/Projects/HSC/HSCProject.html),  EUCLID (http://sci.esa.int/euclid) and LSST (http://www.lsst.org). 
Since the signal of cosmic shear is very small, these surveys plan to observe a huge number of background galaxies to reduce statistical error. 
This means that any systematic errors in the lensing analysis must be controlled to be smaller than the statistic error, roughly saying 1\% $\sim$ 0.1\% error is required for the systematic error.  
For this purpose there have been many methods(Bernstein \& Jarvis 2002; Refregier 2003; Kuijken et al. 2006; Miller et al. 2007; Kitching et al. 2008; Melchior 2011) have been developed and tested with simulation(Heymans et al 2006, Massey et al 2007, Bridle et al 2010 and Kitching et al 2012) .  Although there have been a great progress, it seems that no fully satisfying method is available yet. 

One of the systematic error comes from smearing effect by atmospheric turbulence and imperfect optics. This effect is described by point spread function(PSF) and we need to correct the effect very accurately in order to study the dark energy property . 
Previous approaches of PSF correction adopted some sort of  approximation for the form of PSF which prevents from an accurate correction in some cases.
Recently, we have proposed a new approximation free method of PSF correction called ERA method  (ERA1:Okura and Futamase 2014, ERA2:Okura and Futamase 2015) based on E-HOLICs method(Okura and Futamase 2011, Okura and Futamase 2012, Okura and Futamase 2013) which is a generalization of KSB method(Kaiser at al. 1995) and uses an elliptical weight function to avoid  expansion of weight function in measuring ellipticity. 
The method makes use of the artificial image constructed by re-smearing the observed image to have the same ellipticity with the lensed image. We have confirmed by numerical simulation that the method is free from systematic error, but is restricted to the case that PSF has relatively simple forms. 
In this paper we generalize ERA method to be applicable for realistic situation, namely for complicated shapes of PSF and galaxies. Then we show that the improved ERA method has no systematic error by numerical simulation, and is able to apply for real data of Abell 1689 taken by Subaru Suprime Camera.

This paper is organized as follows.
In section 2, we explain our definitions and notations used in this paper.
In section 3, we explain the idea of PSF correction with idealizing in ERA method.
In section 4, we show simulation tests for this method with several types of galaxy and PSF images, and then we apply the method to real data.
Finally we summarize our method and give some discussion  in section 5. 
\section{The Definitions and The Notations}
In this sections, we introduce the definitions and notations we used in this paper. 
The bold symbol, i.e. angular position on celestial sphere, reduced shear and so on, means complex number.
\subsection{Notations of the brightness distributions}
\label{sec:brightness}
In this paper, many brightness distributions are used for explaining ERA method.
Table \ref{tbl:brightnessdistribution}  surmmarize our notation and definition of the brightness distributions.
The corresponding distributions in Fourier space are written with the same character with hat.
More details of the definition are explained when these are used.
\begin{table}[tbp]
\begin{tabular}{cccl}\hline
object & ID & notation & definition \\
\hline\hline
galaxy & GAL & $G(\bth)$ & Galaxy before smeared by PSF \\
point spread function & PSF & $P(\bth)$ & PSF image measured from star image \\
smeared galaxy & SMD & $S(\bth)$ & Observed galaxy after smeared by PSF \\
deconvolved galaxy & DGAL & $D(\bth)$ & Reconstructed galaxy by deconvolution \\
idealized PSF & IPSF & $I(\bth)$ & Idealized PSF which can be set arbitrary \\
re-smearing function & RSF & $R(\bth)$ & Re-smearing function to make IPSF \\
re-smeared galaxy & RESMD & $S^{re}(\bth)$ & Galaxy after re-smeared by RSF \\
re-smeared PSF & REPSF & $P^{re}(\bth)$ & PSF after re-smeared by RSF \\
\hline
\end{tabular}
\caption{
\label{tbl:brightnessdistribution}
$\bth$ angular position on the celestial sphere.
The brightness distribution $A(\bth)$ is written as $\hatA(\mbk)$ in Fourier space.
}
\end{table}
\subsection{The idea of Zero plane and zero image}
In this section we introduce the idea of the zero plane and the zero image. 
The idea of the zero plane is that the intrinsic ellipticity of the source comes from a (virtual) source with zero ellipticity (called the zero image) in the virtual plane (the zero plane).

\begin{figure*}[htbp]
\epsscale{0.5}
\plotone{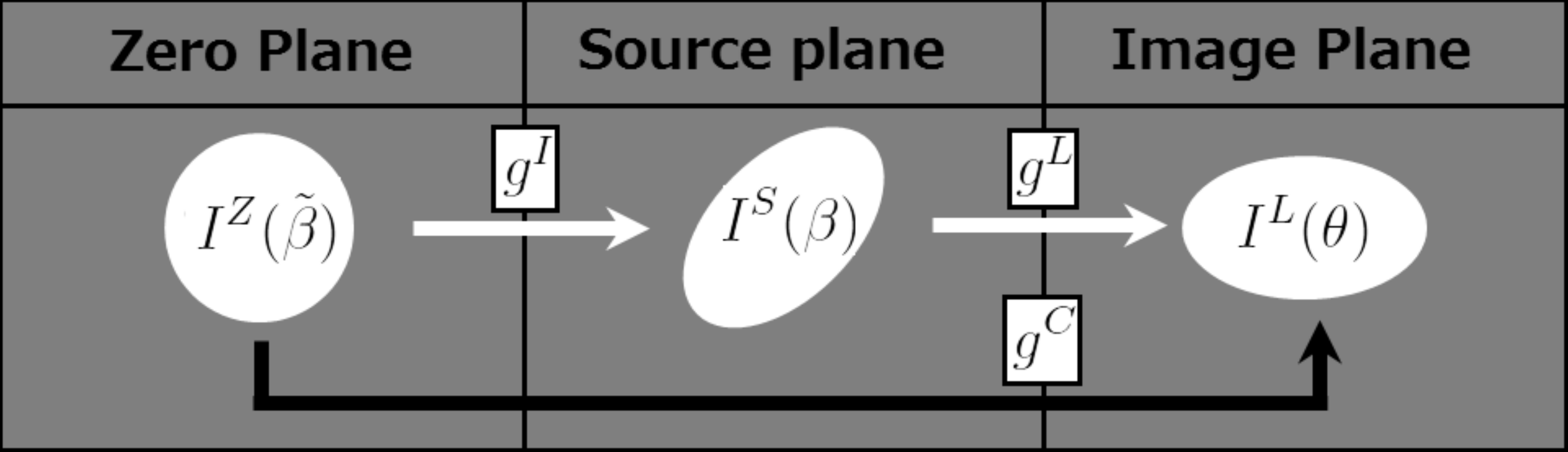}
\caption{
\label{fig:ERA_system0}
The relation between Zero, Source and Image plane.
}
\end{figure*}

Suppose we have the reduced shear due to lensing and the intrinsic ellipticity written as
of lensing and the intrinsic ellipticity respectively as $\mbg^L$ and $\mbg^I$, respectively,  then 
the relationship between the displacements in the zero plane $\tilde \bbe$ and the source plane $\bbe$ and the lens plane $\bth$ are described as
\begin{eqnarray}
\tilde \bbe&=&\bbe -\mbg^I\bbe^*\\
\label{eq:Cshear}
\tilde \bbe&=&\bth -\mbg^C\bth^*,
\end{eqnarray}
where $\mbg^C$ is combined shear which is defined as 
\begin{eqnarray}
\mbg^C \equiv \frac{\mbg^I+\mbg^L}{1+\mbg^I\mbg^{L*}}.
\end{eqnarray}
Figure \ref{fig:ERA_system0} shows the relation between Zero, Source and Image plane.

This combined shear has information of the intrinsic ellipticity and the lensing reduced shear. 
Since the intrinsic ellipticities are random (where we do not consider intrinsic alignment due to galaxy cluster tidal field), the lensing reduced shear can be obtained by removing the information of intrinsic ellipticity as
\begin{eqnarray}
\label{eq:obtainLRS}
\left< \frac{\mbg^C-\mbg^L}{1-\mbg^C\mbg^{L*}} \right>=\left< \mbg^I \right>=0.
\end{eqnarray}

This shows that we can obtain the lensing reduced shear in two steps.
The first step is to obtain the combined shear from each object (Eq.\ref{eq:Cshear}) and the  
second step is to obtain the lensing reduced shear by  averaging (Eq.\ref{eq:obtainLRS}).

In this paper, we consider these two steps to be combined into one,
so we consider only the relationship between the zero plane and the lens plane, 
and we use $\tilde \bbe$ as $\bbe$ and $\mbg^C$ as $\mbg$ for notational simplicity.

\subsection{Ellipticity for the simulation test}
\label{sec:ellipticity}
In this section, we define the ellipticity we used for the following sections.
As we will explain in the section \ref{sec:basics}, the PSF correction in the ERA method does not depend on the definition of the ellipticity for object image, but we need to adopt at least one definition for real analysis and simulation test.

The ellipticity used in this paper is defined by moments of the images.
The complex moments of the measured image are denoted as $\cZ^N_M$ and measured as
\begin{eqnarray}
\label{eq:CMOM}
\cZ^N_M(I,\bep_W)&\equiv&\int d^2\theta \bth^N_M I(\bth) W(\bth,\bep_W)\\
\label{eq:CMOM2}
\bth^N_M&\equiv&\bth^{\frac{N+M}{2}}\bth^{*\frac{N-M}{2}},
\end{eqnarray}
except 0th moments defined as
\begin{eqnarray}
\label{eq:Z02}
\cZ^0_2(I,\bep_W)&\equiv&\int d^2\theta\frac{\bth^2_2}{\bth^2_0}I(\bth) W(\bth,\bep_W)\nonumber\\
&=&\int d^2\theta\lr{\cos (2\phi_\theta)+i\sin (2\phi_\theta)}I(\bth) W(\bth,\bep_W),
\end{eqnarray}
where W is a weight function which is a function of displacement from the centroid $\bth$ and ellipticity $\bep_W$, 
the subscript N is the order of moments and M indicates the spin number, and $\phi_\theta$ is the position angle at $\bth$. 

The ellipticities are defined as spin-2 combination of the quadrupole moments or combination of the 0th moments with normalization.
\begin{eqnarray}
\label{eq:EP2nd}
\bep_{2nd}&\equiv&\lrs{\frac{\cZ^2_2}{\cZ^2_0}}_{\lr{I,\bep_W}}\\
\label{eq:E2nd}
\mbe_{2nd}&\equiv&\frac{\bep_{2nd}}{|\bep_{2nd}|^2}\lr{1-\sqrt{1-|\bep_{2nd}|^2}}\\
\label{eq:E0th}
\mbeZ&\equiv&\lrs{\frac{\cZ^0_2}{\cZ^0_0}}_{(I,\bep_W)}\\
\label{eq:EP0th}
\bepZ&\equiv& 2\mbeZ/(1+|\mbeZ|^2).
\end{eqnarray}
We refer to $\mbeZ$ and $\bepZ$ as the "0th-ellipticity" and $\mbe_{2nd}$ and $\bep_{2nd}$ as the "2nd-ellipticity".
If the profile of the measured image is simple, for example an elliptical Gaussian, the 0th- and 2nd-ellipticities have the same value ($\mbeZ=\mbe_{2nd}$ and $\bep_{0th}=\bep_{2nd}$),
but because a real image has a complex form, these ellipticities usually have different values.
The ellipticity of the weight function $\bep_W$ should be zero in zero plane, so it is $\bep_{2nd}$ or $\bep_{0th}$ in the image plane. By selecting so, we can obtain the ellipticity of the object without any approximation.
The relationship between the ellipticities and the reduced shear is obtained as follows:
\begin{eqnarray}
\mbe=&\mbg&\hspace{50pt}g<1
\\
\mbe=&\frac{1}{\mbg^*}&\hspace{50pt}g>1.
\end{eqnarray}
for $\mbeZ$ and $\mbe_{2nd}$.
More details can be seen in ERA1 and ERA2.

\section{The Basics of PSF correction in the ERA Method}
\label{sec:basics}
In this section we present the basics of the ERA method.

One of the methods of measuring ellipticity of galaxy from smeared image is to reconstruct galaxy image by deconvolution. 
Suppose that (GAL, $G(\bth)$) is the galaxy image without the effect of PSF. This image is smeared by point spread function(PSF, $P(\bth)$), which can be measured from star images, and thus the image ( SMD, $S(\bth)$) we actually observe is related by convolution as
\begin{eqnarray}
\label{eq:GP}
\hatS(\mbk)=\hatG(\mbk)\hatP(\mbk).
\end{eqnarray}
Then,the deconvolved galaxy(DGAL, $D(\bth)$) is defined as 
\begin{eqnarray}
\label{eq:GPS}
\hatD(\mbk)\equiv\frac{\hatS(\mbk)}{\hatP(\mbk)}.
\end{eqnarray}
If there is no noise on the GAL and PSF, the DGAL coincides with GAL, so the brightness distribution of GAL can be obtained as $G(\bth)=D(\bth)$. However, in real observation, the brightness distribution $A(\bth)$ we observed has not only signal 
$A_S(\bth)$ but also noise $A_N(\bth)$, e.g. Poisson noise of sky count, therefore GAL and PSF we observed can be decomposed
\begin{eqnarray}
\label{eq:PS_obs}
P(\bth)&=&P_S(\bth)+P_{N}(\bth)\\
S(\bth)&=&S_S(\bth)+S_{N}(\bth),
\end{eqnarray}
and so DGAL is written as
\begin{eqnarray}
\label{eq:D_obs}
\hatD(\mbk)=\frac{\hatS(\mbk)}{\hatP(\mbk)}=\frac{\hatS_{S}(\mbk)+\hatS_{N}(\mbk)}{\hatP_{S}(\mbk)+\hatP_{N}(\mbk)}
=\frac{\hatG(\mbk)+\hatS_{N}(\mbk)/\hatP_{S}(\mbk)}{1+\hatP_{N}(\mbk)/\hatP_{S}(\mbk)}.
\end{eqnarray}
It is difficult to use DGAL for shear analysis, because DGAL could diverge at some $\mbk$ where $\hatP_{N}(\mbk)\sim-\hatP_{S}(\mbk)$, therefore DGAL does not have an analyzable shape.

The basic idea of the ERA method is resmearing PSF and SMD by re-smearing function (RSF, $R(\bth)$) to idealize the PSF in the following sense.
''Re-smearing'' GAL and PSF by some RSF is defined as 
\begin{eqnarray}
\label{eq:resmearing}
\hatP^{re}(\mbk)&\equiv&\hatP(\mbk)\hatR(\mbk)\\
\hatS^{re}(\mbk)&\equiv&\hatS(\mbk)\hatR(\mbk)=\hatG(\mbk)\hatP(\mbk)\hatR(\mbk)=\hatG(\mbk)\hatP^{re}(\mbk),
\end{eqnarray}
where $P^{re}(\bth)$ and $S^{re}(\bth)$ are re-smeared PSF (REPSF, $P^{re}(\bth)$) and re-smeared galaxy(RESMD, $S^{re}(\bth)$), respectively.
These two equations mean that PSF shape can be chosen arbitrarily because RESMD can be written as convolution of GAL and REPSF, and REPSF is obtained by the re-smearing too. Therefore REPSF is new PSF and the shape can be set arbitrary. 
It is obvious that the simpler PSF is the better for shear analysis, so the idealized PSF (IPSF, $I(\bth)$) should be defined to be an elliptical simple function, e.g. simple elliptical Gaussian, then RSF is obtained as
\begin{eqnarray}
\label{eq:RSF}
\hatR(\mbk)&\equiv&\frac{\hatI(\mbk)}{\hatP(\mbk)},
\end{eqnarray}
where we must be careful with the divergence, but it can be avoided by noticing the arbitrariness of the IPSF. We discuss about this point later.
Finally, PSF can be idealized as
\begin{eqnarray}
\label{eq:idealizing}
\hatP^{re}(\mbk)&=&\hatI(\mbk)\\
\hatS^{re}(\mbk)&=&\hatG(\mbk)\hatI(\mbk).
\end{eqnarray}

Next, we explain how to choose IPSF for getting ellipticity of GAL in a simple case. Let GAL have one ellipticity, that is the ellipticity without radial dependence. Then IPSF is a simple function and has same ellipticity with GAL, and the RESMD has also the same ellipticity with GAL, because the three images can be made by affine transformation with same shear from three circular images. Therefore IPSF will be chosen in such a way that it makes RESMD to have the same ellipticity with GAL. In this re-smearing, we do not mention about size of IPSF. It is a free parameter in our case because we need only the ellipticity of GAL, so the ellipticity of RESMD, and sizes of GAL and RESMD are not important in this simple situation. One of the important thing is that the way how to define the ellipticity is irrelevant in our method. We can use any definition of ellipticity if we know the relation between the ellipticity and gravitational shear. More details can be seen in ERA1.

In general case, GAL does not have simple shape and has radially dependent ellipticity, therefore we need to consider the situation that the re-smearing procedure averages over the radial depended ellipticity.
 The ellipticity of the observed galaxies consists of the intrinsic ellipticity of galaxy and gravitational shear. The intrinsic ellipticity has a radial dependence in general but shear has no radial dependence, so the re-smearing merges only intrinsic ellipticity. This means that the intrinsic ellipticity has different value by different re-smearing size. Reasonable choice for the size of IPSF would be PSF size. However this choice might cause the divergence we mentioned above. One of the technique to avoid the divergence is to use a re-smearing function larger in size than PSF. Therefore it seems natural to use IPSF which has a size as small as possible but larger than PSF. Then the problem now is that the size depends on PSF of each exposures, and makes different intrinsic ellipticity in each exposures. According to the above consideration, we choose a fixed size of IPSF for multi epoch exposures which is slightly larger than the maximum PSF size of all exposures. It will obviously depend on optics, seeing and so on.

The maximum value $\mu_{max}$ of the ratio $\mu$ between IPSF and PSF in Fourier space will be used to check the divergence
\begin{eqnarray}
\label{eq:size_fit}
\mu=\lr{\frac{\hatI(\mbk)}{\hatI(\mbk_0)}}/\lr{\frac{\hatP(\mbk)}{\hatP(\mbk_0)}}
\end{eqnarray}
in all $\mbk$ except $\mbk_0$, where $\mbk_0$ is a position of peak of the function.

If some objects have $\mu_{max}$ much larger than 1, it may be due to pixel noise or some other noise, we should remove these objects in the shot. But, if many objects have μmax much larger than 1, it is probably due to bad choice of the size for IPSF. The choice depends on how large the noise the object images have. This situation will be studied in detail in future works. In the following simulation test, we use the size of IPSF 1.5 times larger than the measured Gaussian best size. Then we test how does precision in ellipticity measurement change with the IPSF size.

In the above we discussed that the size of IPSF is not directly related with PSF size. However, the size should be depend directly on ellipticity of GAL.
The lensing shear is one type of affine(linear) transformations, and the transformation relates shapes between with/without shear distortion.
This is very useful, especially in simulation test, because we can obtain the same intrinsic elipticity from images with/without lensing shear.
Let's consider the IPSF $I^0(\tilde \bbe)$ in the zero plane. In the zero plane, GAL has 0 ellipticity by the definition, therefore the IPSF is a circular function. The IPSF $I(\bth)$ in the lens plane should have a profile which is transformed by shear distortion from $I^0(\tilde \bbe)$, so 
\begin{eqnarray}
\label{eq:s_IPSF}
I(\bth)=I^0(\tilde \bbe).
\end{eqnarray}
For examples, if we use an elliptical Gaussian for IPSF, then it is described in the zero plane as 
\begin{eqnarray}
I^0(\tilde \bbe)={\rm exp}\lr{-\frac{\bbe^2_0}{\sigma^2}},
\end{eqnarray}
the IPSF in lens plane should be 
\begin{eqnarray}
I(\bth)&=&{\rm exp}\lr{-\frac{\bth^2_0-\bth^2_2\cdot\bde}{\lr{1+|\mbg|^2}\sigma^2}}\\
\bde&\equiv&\frac{2\mbg}{1+|\mbg|^2}.
\end{eqnarray}
So in this situation, by using IPSF  with ellipticity dependent size with factor $\sqrt{1+|\mbg|^2}$, we can obtain the same intrinsic ellipticity in PSF correction for different lensing shear. We use this IPSF size determination in the following simulation test.

One of the method to determine the ellipticity of IPSF is to use the iteration like
\begin{eqnarray}
\label{eq:e_IPSF}
e_{IPSF}^{i+1}=e_{IPSF}^{i}+e_{RESMD}^{i}-e_{REPSF}^{i}
\end{eqnarray}
for $i$th iteration, where the ellipticity of SMD can be used for initial ellipticity $e_{IPSF}^{0}$.

\section{simulation test}
\label{sec:test_all}
In this section, we present simulation test of precision and analysis speed in the shear measurement by ERA method. In this test, we analyze the ellipticity or lensing shear from several simulated objects with different situations to investigate what makes systematic error. The profile of IPSF is an elliptical Gaussian, the size is 1.5 times larger than the size of the corresponding PSF as Gaussian. Ellipticity of images are defined by quadrupole moment(2nd ellipticity) and 0th moment(0th ellipticity). Then we also test the time required for analyze objects which have realistic size. All simulated images have no pixel noise to investigate the systematic error only from the PSF correction. The systematic error from pixel noise will be investigated in future works.

\subsection{Test with several PSF model}
\label{sec:test_psf}
In this section we consider the most simple situation where galaxies and PSFs have large size enough to ignore pixelization. Then galaxies has only one profile (Gaussian or Sersic) and so one ellipticity, but we consider several types of PSF with complex shapes. Although the shapes of GAL and PSF in this test is not realistic, it is useful to investigate systematic error only from PSF shape.

\begin{figure*}[htbp]
\centering
\resizebox{1.0\hsize}{!}{\includegraphics{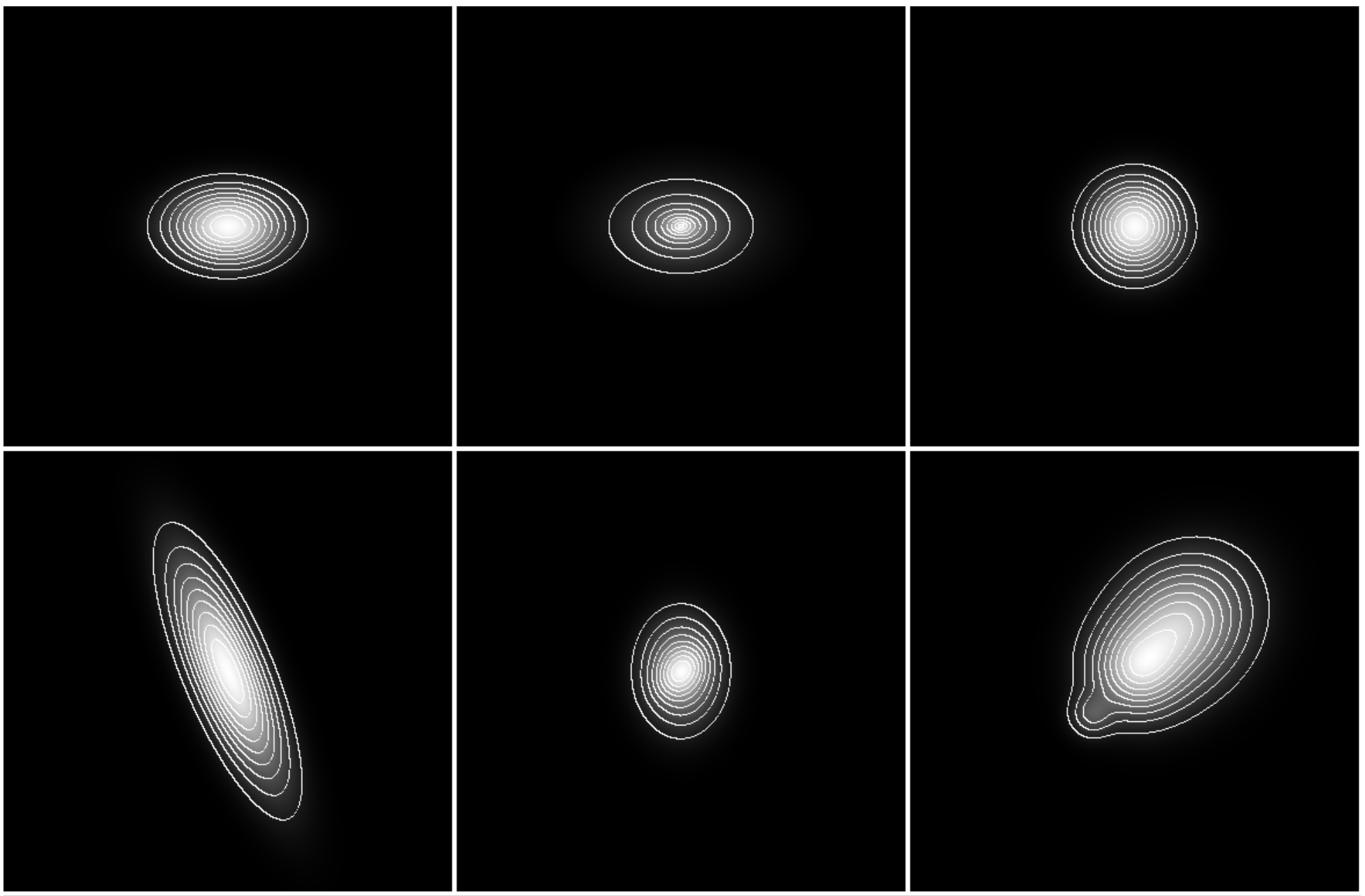}}
\caption{
\label{fig:psfs}
Simulated images of galaxies and PSFs, these are shown with same size scale.
Object type ID = [G, S, C] from left to right in top panel. 
Object type ID = [E, D, T] from left to right  in bottom panel. 
}
\end{figure*}

We use Gaussian (Type = G) and Sersic (Type = S) profile for the galaxy images with ellipticity = [0.4, 0.0], this ellipticity is the true ellipticity in this simulation, and 100 pixels Gaussian radius for Type G and corresponded Gaussian size for Type S. Then we assumed 4 types of PSF, the first is a circular Gaussian (Type = C), the second is a highly elliptical Gaussian (Type = E), the third is a double Gaussians(Type = D), the forth is a combination of three Gaussians with position shift (Type = T). Table \ref{tbl:Summay_sim_image} shows the parameters of the simulated images and figure  \ref{fig:psfs} show the simulated images with linear contour.

The precision is quantified by systematic error $\Delta \bep$ which is defined as 
\begin{eqnarray}
\label{eq:syserror}
\Delta \bep_{X} \equiv \frac{\bep^{corrected}_{X}-\bep^{true}}{\bep^{true}},
\end{eqnarray}
where $\bep^{corrected}$ is PSF corrected ellipticity we use for measuring shear and $\bep^{true}$ is true ellipticity of galaxy, X is ``2nd'' for 2nd ellipticity or ``0th' for 0th ellipticity'.
Table \ref{tbl:Result_sim_psf} shows results of the precision test. 
The systematic errors in the PSF corrections are smaller than 0.1\%. Therefore one can conclude that the systematic error in PSF correction for simple shape of galaxy can be ignored even if the PSF has complicated shape.

\begin{table}[tbp]
\begin{tabular}{ccccc}\hline
Object  &Type ID &Profile & ellipticity & Gaussian radius 
\\\hline\hline
Galaxy & G & Gaussian & 0.4, 0.0 & 100 pixels\\\hline
Galaxy & S & Sersic      & 0.4, 0.0 & 100 pixels$^{1)}$\\\hline
PSF     & C & Gaussian & 0.0, 0.0 & 100 pixels\\\hline
PSF     & E & Gaussian &-0.6,-0.6 & 100 pixels\\\hline
PSF     & D & Gaussian & 0.0, 0.3 &   50 pixels\\
            &    &+ Gaussian &-0.3, 0.0 & 100 pixels\\\hline
PSF     & T & Gaussian & 0.0, 0.0 &  100 pixels\\
            &    & +Gaussian$^{2)}$ &0.0, 0.2 & 150 pixels\\
            &    & +Gaussian$^{3)}$ & 0.0, 0.0 &   50 pixels\\\hline
\end{tabular}
\caption{
\label{tbl:Summay_sim_image}
${1)}$ Corresponded size as Gaussian size.
${2)}$ with position shift [+150, +150] pixels. 
${3)}$ with position shift [-150, -150] pixels.
}
\end{table}
\begin{table}[tbp]
\begin{tabular}{cccccc}\hline 
Galaxy ID &Type PSF & $r_{PSF}$[pixels] & $r_{IPSF}$[pixels] & $\Delta \bep_{2nd}$ & $\Delta \bep_{0th}$ 
\\\hline\hline
G & C & 100.0 & 150.0 & 1.343e-7 & 8.764e-8 \\
G & E & 100.0 & 350.0 &-2.425e-6 &-2.349e-6 \\
G & D &  90.2  & 150.0 &-2.973e-7 &-4.256e-7 \\
G & T & 150.0 & 275.0 &-3.260e-6 &-9.045e-6 \\
S & C & 100.0 & 150.0 &-6.765e-8 & 5.201e-7 \\
S & E & 100.0 & 350.0 & 2.483e-8 &-3.050e-7 \\
S & D &  90.2  & 150.0 & 5.747e-7 &-3.376e-7 \\
S & T  & 150.0 & 275.0 &-8.324e-7 &-1.757e-6 \\\hline
\end{tabular}
\caption{
\label{tbl:Result_sim_psf}
The systematic  error in shear measurement.
The radius of PSF is best radius of weight function for measuring PSF shape with elliptical Gaussian weight.
}
\end{table}

\subsection{Test with pixelized galaxy}
\label{sec:test_pixel}
Next, we test systematic error and analysis speed for realistic size of pixelized image. We selected SMD(ID = G) and PSF(ID = C) which are used in section \ref{sec:test_psf}. We rescale them to GAL which has Gaussian size 2.0 or 4.0 pixels. This means that SMD size and PSF size are 50 times and 25 times scale down with pixelization. Table \ref{tbl:Result_sim_pixelized} shows the systematic error $\Delta \bep$ of the pixelized images.

We can see that the measurement by using 0th ellipticity has systematic error of the order of 0.1\% compared with very small errors by 2nd ellipticity. We guess that the larger systematic error comes from center and surround pixels, because in such region brightness distribution fluctuates rapidly over pixel scale. This does not necessary mean that 0th ellipticity can not be used for cosmic shear measurement, because 0th ellipticity has higher, approximately 1.5 times, signal to noise ratio than 2nd ellipticity, so 0th ellipticity has less systematic error from pixel noise, and so it may be useful for shear measurement from faint galaxies. One of the idea which solves this systematic error form pixelization is to reduce weight at center and surround pixels. We will study about the weight and other method to correct the systematic error in the forthcoming publication.

Table \ref{tbl:Result_sim_pixelized} also shows the analysis speed $T$ which is average time over the time analyzing 1000 same images. 
We can see that the analysis speed depends on the radius of objects, but in any cases, the speed is about 0.1 seconds or shorter.
This speed is for simulated images, so it is expected that the speed becomes longer for real object images due to pixel noise and size variation of galaxies. We test the analysis speed time for real objects in section \ref{sec:test_real}.

\begin{table}[tbp]
\begin{tabular}{ccccccc}\hline
radius of GAL & $r_{PSF}$[pixels] & $r_{IPSF}$[pixels] & $\Delta \bep_{2nd}$ & $\Delta \bep_{0th}$ & $T_{2nd}$ &  $T_{0th}$ \\
\hline\hline
2.0 pixels & 2.0 & 3.0 & 9.580e-7 &-2.221e-3 & 0.024 & 0.028 \\
4.0 pixels & 4.0 & 6.0 & 9.705e-7 &-1.236e-4 & 0.072 & 0.119 \\\hline
\end{tabular}
\caption{
\label{tbl:Result_sim_pixelized}
The unit of the analysis speed is seconds/object.
}
\end{table}

\subsection{Test with double elliptical galaxy and spiral galaxy}
\label{sec:test_gal}
Next, we test the systematic error in the measurement for images with radially dependent ellipticity. Since the ellipticity changes with radius, we can not define an unique ellipticity for such image. In order to test the precision of the measurement for such images, we consider four galaxy images with four different intrinsic ellipticities whose directions are 90 degree different each other and distort them by the same amount of shear $\gamma^{true}=[0.1, 0.0]$. 
Then we measure the shear $\gamma^{measured}_X$ by averaging the combined shear of these four images, then we define the systematic error ratio $\Delta\gamma_X$, X='GAL'(no PSF) or '2nd' or '0th', as
\begin{eqnarray}
\label{eq:syserror_gamma}
\Delta \gamma_{X} \equiv \frac{\gamma^{measured}_{X}-\gamma^{true}}{\gamma^{true}},
\end{eqnarray}

\begin{figure*}[tbp]
\centering
\resizebox{1.0\hsize}{!}{\includegraphics{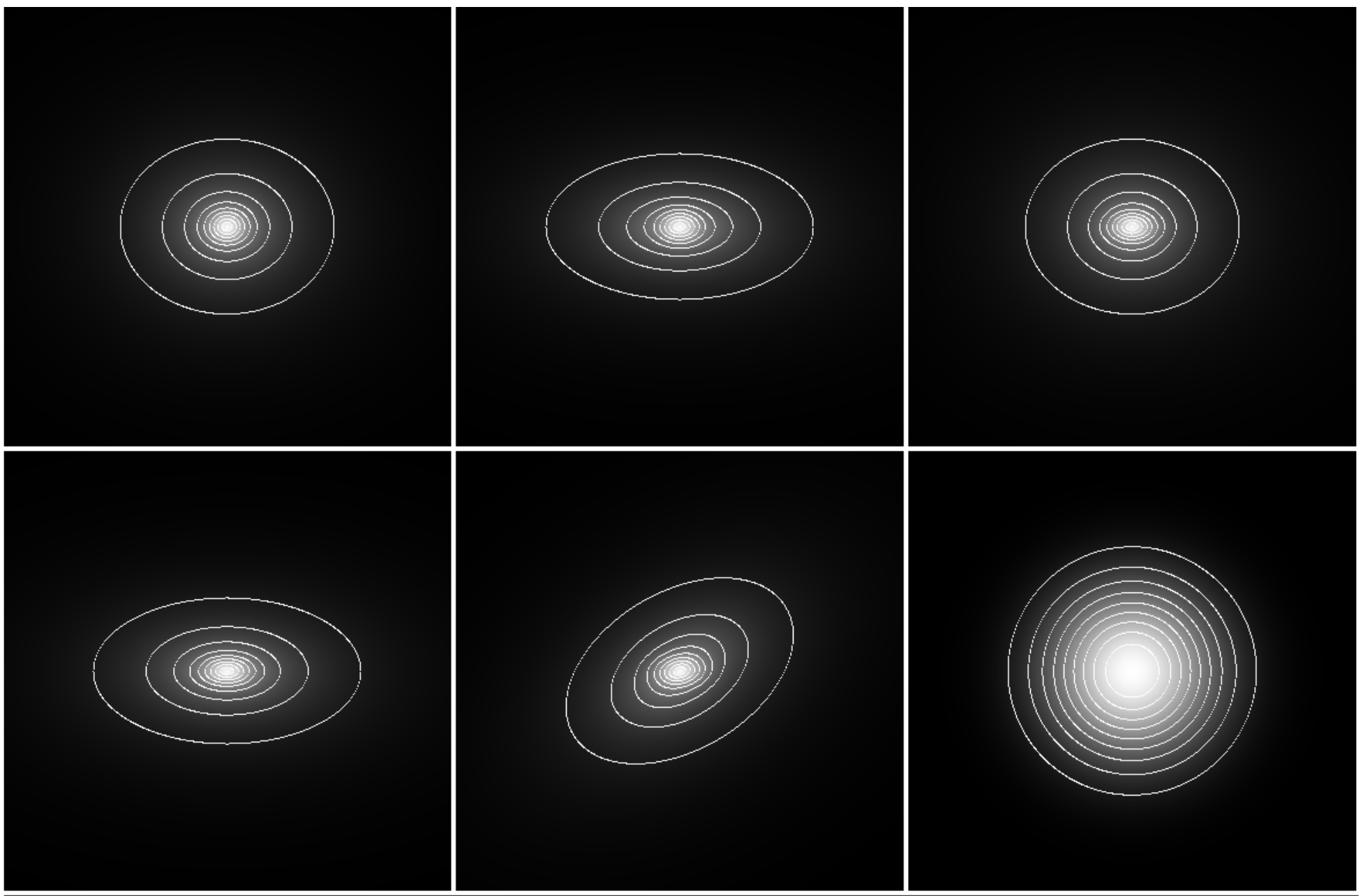}}
\caption{
\label{fig:doubles}
Simulated images of galaxies(original rotation) and psf with shear distortion, these are shown with same size scale.
ID = [00, 02, 20] from left to right in top panel. 
ID = [22, 2R2, psf] from left to right  in bottom panel. 
}
\end{figure*}

\begin{figure*}[tbp]
\centering
\resizebox{1.0\hsize}{!}{\includegraphics{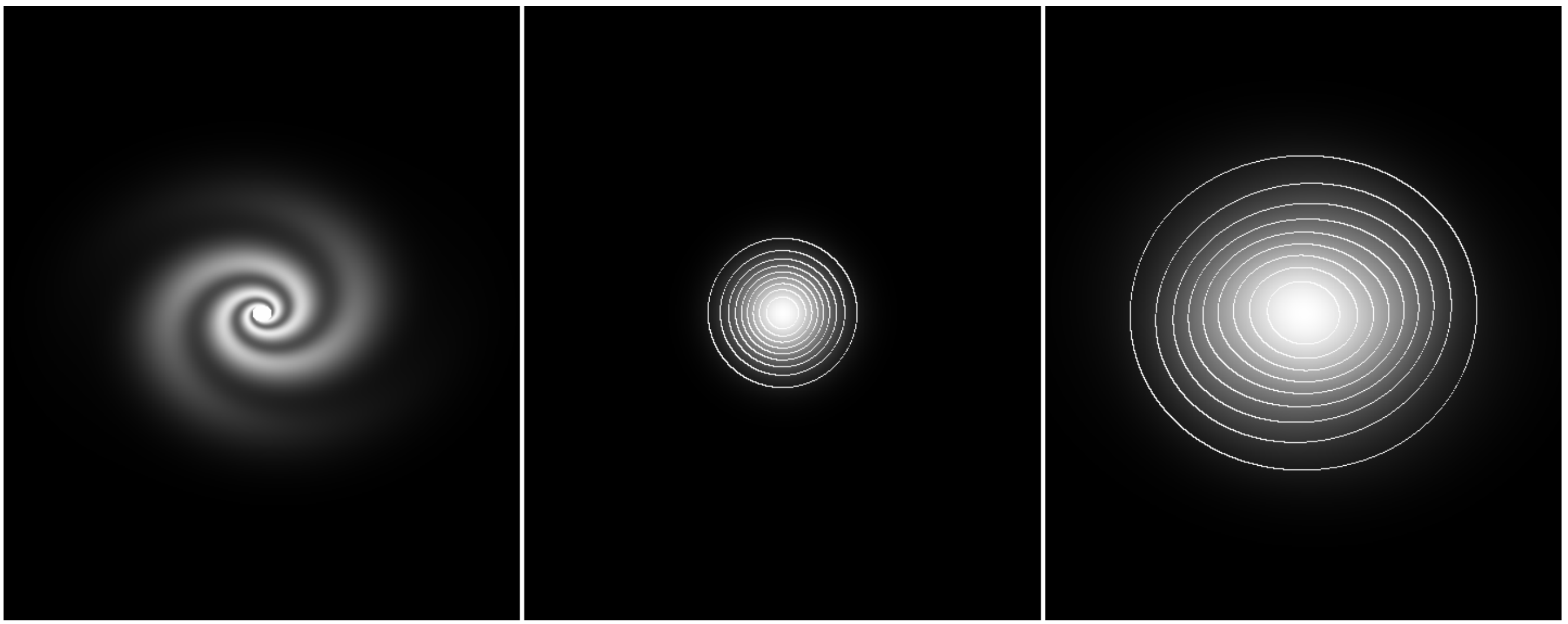}}
\caption{
\label{fig:spiral}
Simulated images of spiral galaxies(left), PSF(center) and smeared galaxy(right), with same size scale.
The smeared galaxy has radial depended ellipticity.
}
\end{figure*}

First, we test with 5 data sets of galaxies which have two ellipticities with Gaussian profile core and Sersic profile tail.
The Gaussian core has 10 pixels Gaussian radius and the Sersic tail has 50 pixels radius as the corresponding Gaussian radius, then  PSF is a circular Gaussian with $r_{PSF}=50$ pixels and IPSF has an elliptical Gaussian with $r_{IPSF}=75$ pixels.
Figure \ref{fig:doubles} shows the original direction images of each the 5 data sets and PSF with the same size scale.
Each image sets has ID=[00, 02, 20, 22, 2R2], where the first(second) number means intrinsic shear of Gaussian(Sersic) and ``0''(``2'') means 0.0(0.2), ``R'' means that the direction of ellipticity of the Sersic rotates 45 degree from the direction of ellipticity of the Gaussian profile.
Next, we test galaxy with two spiral arms. Figure \ref{fig:spiral} shows the simulated spiral galaxy image, PSF image and the smeared spiral galaxy image with same size scale. We can see that the smeared galaxy image has radial dependent ellipticity.
In this test, PSF has a circular Gaussian profile with $r_{PSF}=64$ pixels, so we set $r_{IPSF}=96$ pixels for IPSF.
Figure \ref{tbl:Result_sim_double} shows the systematic error in the shear measurement.
We can see that the systematic error is smaller than $0.1\%$ in all cases, this means PSF effect can be corrected precisely enough even if the GAL has radial dependent ellipticity.

\begin{figure*}[tbp]
\centering
\resizebox{1.0\hsize}{!}{\includegraphics{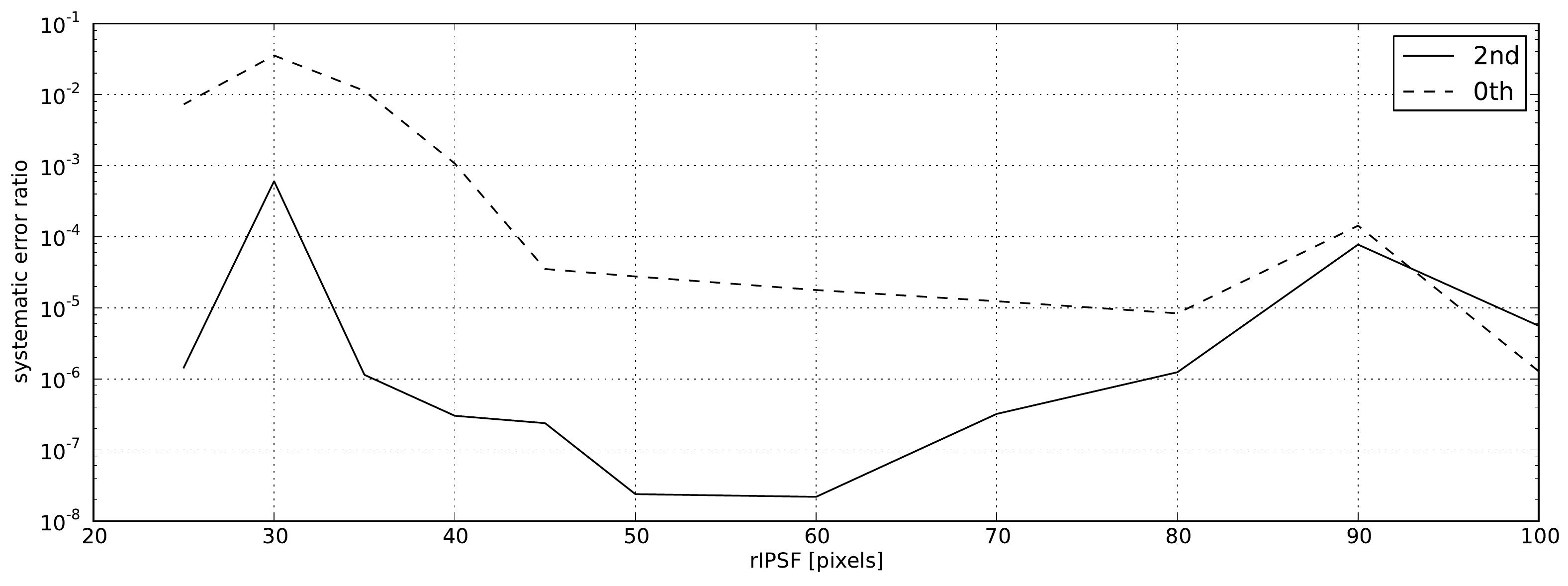}}
\caption{
\label{fig:rIPSF}
The systematic error ratio by each $r_{IPSF}$ with PSF size = 50 pixels. The (dashed )line is the error ratio in 2nd(0th) ellipticity.
 }
\end{figure*}

Then, we test the size dependence of IPSF for the systematic error. We measured shear in above situation ID = ``20''  except that IPSF size which is selected in the range between half and twice of PSF size. As figure \ref{fig:rIPSF} shows,  the result is precise enough precision if IPSF size is selected from higher than PSF size.  
This test is done under the ideal situation, so the result may change in more realistic situation, e.g. with pixel noise. However it is important to test intrinsical systematic error in selecting size of IPSF.
The test with pixel noise will be studied in future.

Finally we test the situation with different PSF shapes. In this test, we use four galaxy images with ID = ``02'' with different rotation same as above test and distorted by the same shear [0.1, 0.0], and these four galaxy images are smeared by four PSFs, these images has $50$ pixels size but different shapes. One of the PSFs has ID = ``C'', so it is circular, and others has ID = ``D'' with different directions. IPSF is an elliptical Gaussian with fixed size $75$ pixels.
The last row of the table \ref{tbl:Result_sim_double} shows the result, and the systematic error ratio is smaller than $0.1\%$. This means that the PSF correction has enough precision for galaxy with radially dependent ellipticity and PSF variances.

\begin{table}[tbp]
\begin{tabular}{ccccccc}\hline
Galaxy ID  & $\Delta \gamma_{GAL}$ & $\Delta \gamma_{2nd}$ & $\Delta \gamma_{0th}$
\\\hline\hline
00      & 3.254e-8 &-7.225e-7 & 1.029e-5 \\
02      & 1.455e-8 &-5.661e-5 &-1.067e-4 \\
20      &-1.230e-7 &-7.182e-7 & 1.031e-5 \\
22      &-9.077e-9 &-8.600e-7 &-6.371e-6 \\
2R2   & 3.087e-7 &-4.277e-7 & 1.115e-5 \\
spiral &-6.286e-6 &-8.332e-5 &-1.259e-4 \\
multi  &-1.536e-8 &-3.387e-5 &-5.264e-5 \\
\hline\end{tabular}
\caption{
\label{tbl:Result_sim_double}
The systematic error in the shear measurement from the double elliptical Galaxies.
}
\end{table}

\subsection{Test with real data}
\begin{figure*}[tbp]
\centering
\resizebox{1.0\hsize}{!}{\includegraphics{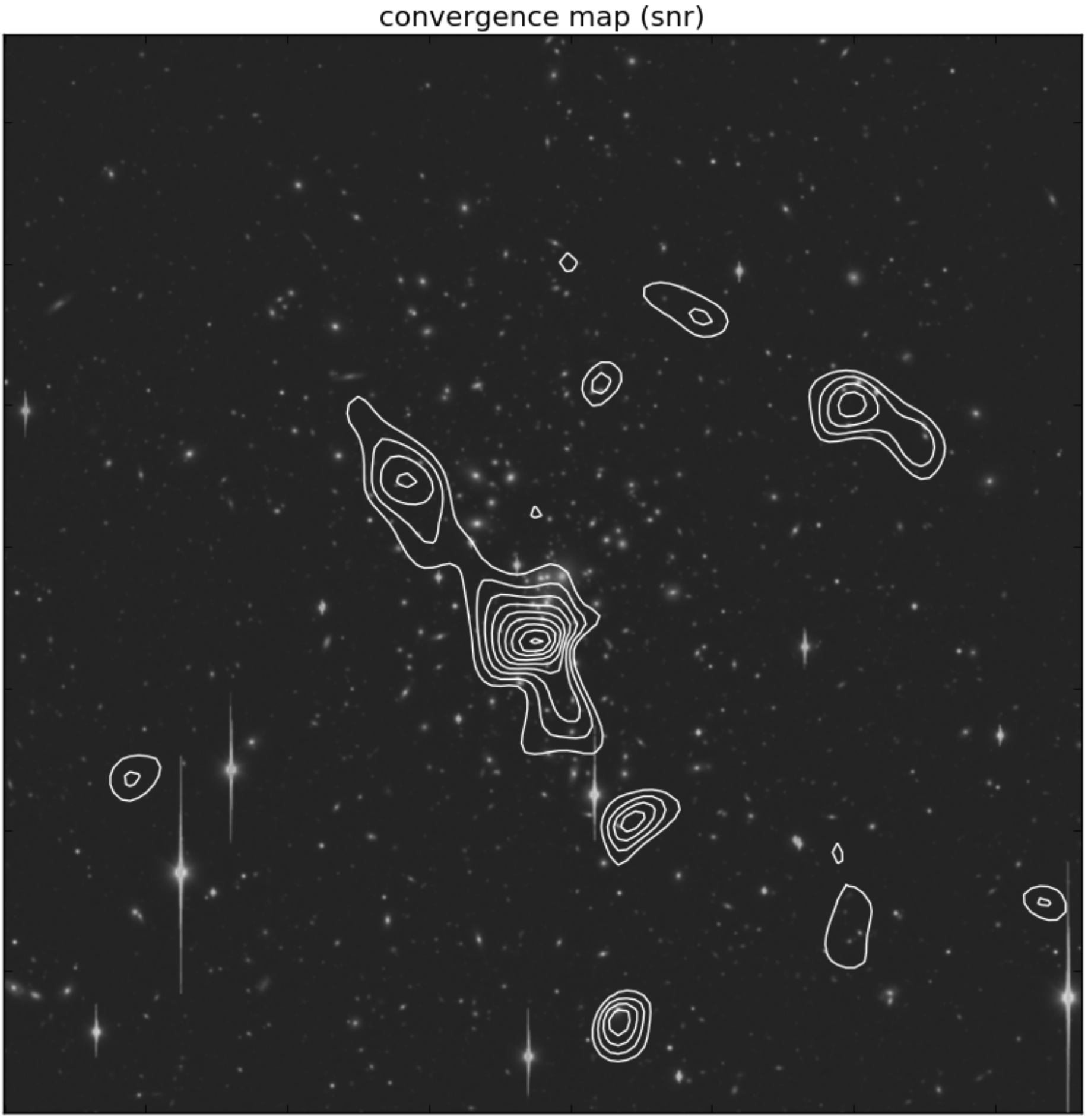}}
\caption{
\label{fig:A1689}
Reconstructed convergence of Abell 1689 galaxy cluster.
The steps of the contours means 0.5 signal to noise ratio and the lowest contour means 2.0.
}
\end{figure*}

Now we apply our method of PSF correction for real data and test analysis speed with real observed image. We used 3000 $\times$ 3000 square pixels image including A1689 galaxy cluster, then approximately 1500 objects were selected as galaxy.
The analysis time for each objects can be seen in table \ref{tbl:Result_real_A1689}.
The 4th column of the table is total time over the succeeded number, so it includes time for rejection for the rejected objects.
The 3rd column of the table is mean time of analysis time of only objects succeeded in PSF correction, so it does not include time for the rejection, this is meaningful as re-analyzing time. 
The mean size of galaxies we succeeded in analyzing shear is 2.46 pixels. 
The time required for analyzing shear from real data is roughly 2 or 3 times longer than simulated images with 2.0 radius images.
We think this is reasonable time, because there are the radius distribution and non-simple shape of the galaxies and pixel noise on the images and so on in real data.
As the last test, we reconstructed the weak lensing convergence map of Abell 1689 galaxy cluster.
The aim of this test is to check if the ERA method can obtain the systematic shear pattern, and we do not intend to determine the mass distribution of the cluster very accurately. For this aim, it is enough to select the back ground galaxies simply from their sizes and magnitude and signal to noise ratio.
The image data has 100 arcmin$^2$ size and approximately 7 selected back ground galaxies per 1 arcmin$^2$.
The figure \ref{fig:A1689} shows the reconstructed convergence, the steps of the contours means 0.5 signal to noise ratio and the lowest contour means 2.0.
We can see that the convergence has a peak in the center of the cluster as expected. Thus the ERA method can be applied to real data and be successfully obtained the systematic weak lensing shear pattern.
\label{sec:test_real}
\begin{table}[tbp]
\begin{tabular}{ccccc}\hline
 & selected number  & number analyzed   & $T^{all}_{X}$  & $T^{succeeded}_{X}$ \\
\hline\hline
2nd & 2180 & 2163 & 0.0499& 0.0475\\
0th &  2144 &  2143 & 0.0924 & 0.0906\\\hline
\end{tabular}
\caption{
\label{tbl:Result_real_A1689}
The unit of the analysis speed is seconds/object.
}
\end{table}

\section{Summary}
We have previously developed a new method of PSF correction in weak gravitational shear analysis called ERA method without adopting any approximation to PSF. In this paper we improve the ERA method by using an ideal PSF in order to treat more realistic PSF and galaxy shape. The correction re-smears PSF and SMD again by RSF for measuring PSF corrected ellipticity with the following steps
\begin{itemize}
\label{eq:REQ}
\item Determine profile of IPSF $I(\bth)$, e.g. an elliptical Gaussian function.
\item Decide the size of IPSF by considering PSF sizes in multi exposures, and it must be fixed in the analysis.
\item Calculating RSF $R(r_{IPSF}, e_{IPSF} ; \bth)$ from IPSF with a trial ellipticity $e_{IPSF}$ and PSF.
\item Re-smear the SMD and the PSF by the RSF .
\item Compare the ellipticity of RESMD and REPSF. If the differences of the two ellipticity is small enough, it is the ellipticity of GAL. If the differences is larger than the precision you want, try re-smearing again with modifying ellipticity of IPSF.
\end{itemize}

We tested the ERA method with several types of simulated galaxies and star images. We consider not only simple shapes such as a simple Gaussian and Sersic, but also more complicated shapes, such as galaxy with radially dependent ellipticity, spiral galaxy, highly elliptical PSF, PSF with pointing error. The results show that our method is able to correct PSF with systematic error under than 0.1\% in any situation we consider. We also applied the ERA method to real data with Abell 1689 cluster to check the analysis speed and to confirm that the systematic shear pattern can be obtained. It turned out that the ERA method can analyze real objects faster than 0.1 seconds for 2nd ellipticity and faster than 1 second for 0th ellipticity.

In this paper, we have not considered systematic error caused by pixelization effect and pixel noise. For more realistic treatment and accurate estimation of cosmic shear, we cannot neglect these effects and we will come back these study in future.
\acknowledgements
This work is supported in part by a Grant-in-Aid for Scientific Research from JSPS (No. 26400264 for T.F.).

\end{document}